\documentclass[submission,copyright]{eptcs}
\providecommand{\event}{INFINITY 2010}

\usepackage[latin1]{inputenc}
\usepackage{amssymb}
\usepackage[pdftex]{graphicx,color}

\newcommand{\reals}{\mathbb{R}}
\newcommand{\ints}{\mathbb{Z}}

\newcommand{\nats}{\mathbb{N}}

\title{Implicit Real Vector Automata\thanks{This work is supported by
the {\em Interuniversity Attraction Poles\/} program {\em MoVES\/} of
the Belgian Federal Science Policy Office, and by the grant 2.4530.02
of the Belgian Fund for Scientific Research (F.R.S.-FNRS).}}

\author{Bernard Boigelot \qquad Julien Brusten\thanks{Research
fellow (``Aspirant'') of the Belgian Fund for Scientific Research
(F.R.S.-FNRS).} \qquad Jean-François Degbomont
\institute{Institut Montefiore, B28\\
Université de Liège\\
B-4000 Liège, Belgium}
\email{\{boigelot,brusten,degbomont\}@montefiore.ulg.ac.be}}

\def\titlerunning{Implicit Real Vector Automata}
\def\authorrunning{B. Boigelot, J. Brusten, J.-F. Degbomont}

\newtheorem{theorem}{Theorem}
\newtheorem{property}{Property}
\newtheorem{definition}{Definition}
\newcommand{\proof}{\mbox{\it Proof:\/}}
\newcommand{\sketch}{\mbox{\it Proof sketch:\/}}
\newcommand{\qed}{\hfill$\Box$}

\begin{document}

\maketitle

\begin{abstract}
This paper addresses the symbolic representation of non-convex real
polyhedra, i.e., sets of real vectors satisfying arbitrary Boolean
combinations of linear constraints. We develop an original data
structure for representing such sets, based on an implicit and concise
encoding of a known structure, the Real Vector Automaton.  The
resulting formalism provides a canonical representation of polyhedra,
is closed under Boolean operators, and admits an efficient decision
procedure for testing the membership of a vector.
\end{abstract}

\section{Introduction}

Algorithms and data structures for handling systems of linear
constraints are extensively used in many areas of computer science
such as computational geometry~\cite{GO04}, optimization
theory~\cite{Sch86}, computer-aided
verification~\cite{CH78,HRP94}, and constraint
programming~\cite{RvBW06}. In this paper, we consider 
systems defined by arbitrary finite Boolean combinations of linear 
constraints over real vectors. Intuitively, a non-trivial linear 
constraint in the $n$-dimensional space describes either a $(n-1)$-plane, 
or a half-space bounded by such a plane. A Boolean combination of
constraints thus defines a region of space delimited by planar
boundaries, that is, a {\em polyhedron\/} (also called {\em
  $n$-polytope\/}).

Our goal is to develop an efficient data structure for representing
arbitrary polyhedra, as well as associated manipulation
algorithms. Among the requirements, one should be able to build
representations of elementary polyhedra (such as the set of solutions
of individual constraints), to apply Boolean operators in order to
combine polyhedra, and to test their equality, inclusion,
emptiness, and whether a given point belongs or not to a polyhedron.

A typical application consists in representing objects in a 3D
modeling tool, in which shapes are approximated by polyhedral
meshes. By applying Boolean operators, the user can modify an object,
for instance, drilling a circular hole amounts to computing the
Boolean difference between the object and a polyhedron approximating a
cylinder. This application requires an efficient implementation of
Boolean operations: A local modification performed on a complex
object should ideally only affect a small part of its representation.

Another application (actually our primary motivation for studying this
problem) is the symbolic representation of the reachable data values
computed during the state-space exploration of programs. In this
setting, a reachable set is computed iteratively, by repeatedly adding
new sets of values to an initial set, and termination is detected by
checking that the result of an exploration step is included in the set
of values that have already been obtained. In this application, it is
highly desirable for a representation of a set to be independent from
the history of its construction, since reachable sets often have
simple structures, but are computed as the result of long sequences of
operations. We are particularly interested in {\em linear hybrid
  systems\/}~\cite{AHH93}, for which symbolic state-space exploration
algorithms have been developed~\cite{ACH+95,Hen96}, requiring efficient data
structures for representing and manipulating systems of linear
constraints. Existing representations either fail to be
canonical~\cite{HRP94,HH04}, or impose undue restrictions on the linear
constraints that can be handled~\cite{Dil89}.

For some restricted classes of systems of linear constraints, data
structures with good properties are already well known.  Consider
for instance conjunctions of linear constraints, which correspond
to convex polyhedra. A convex polyhedron can indifferently be
represented by a list of its bounding constraints, or by a finite set
of vectors (its so-called vertices and extremal rays) that precisely
characterize its shape~\cite{MRTT53}. An efficiently manageable
representation is obtained by combining the bounding constraints and
the vertices and rays of a polyhedron into a single
structure~\cite{CH78,LeV92,BRZH02}.

There are several ways of obtaining a representation suited for
arbitrary combinations of linear constraints. A first one is to
represent a set by a logical formula in additive real arithmetic.
This approach is not efficient enough for our intended applications,
since testing set emptiness, equality, or inclusion become NP-hard
problems~\cite{FR75}. A second strategy is to decompose a non-convex
polyhedron into an explicit union of convex polyhedra (which may
optionally be required to be pairwise disjoint).  The main
disadvantage of this method is that a set can generally be decomposed
in several different ways, and that checking whether two
decompositions correspond to the same set is costly. Moreover,
simplifying a long list of convex polyhedra into an equivalent shorter
union is a difficult operation.

Another solution is to use automata~\cite{BJW05}. The idea is to
encode $n$-di\-men\-sional vectors as words over a given alphabet, and to
represent a set of vectors by a finite-state machine that accepts the
language of their encodings. This technique presents several
advantages.  First, with some precautions, computing Boolean
combinations of sets reduces to applying the same operators to the
languages accepted by the automata that represent them, which is
algorithmically simple. Second, provided that one employs
deterministic automata, checking whether a given vector belongs to a
set becomes very efficient, since it amounts to following a single
path in a transition graph. Finally, some classes of automata can
easily be minimized into a canonical form. This approach has already
been applied successfully to the representation of arbitrary
combinations of linear constraints, yielding a data structure known as
the {\em Real Vector Automaton (RVA)\/}~\cite{BBR97,BJW05}.

Even though RVA provide a canonical representation of polyhedra, and
admit efficient algorithms for applying Boolean operators, they also
have major drawbacks. First, they cannot handle efficiently linear
constraints with coefficients that are not restricted to small
values, since the size of RVA generally gets proportional to the
product of the absolute values of these coefficients~\cite{BRW98}. Second, RVA
representing subsets of the $n$-dimensional space get unnecessarily
large for large values of $n$.

The contribution of this paper is to tackle the first drawback. We
introduce a data structure, the {\em Implicit Real Vector Automaton
  (IRVA)\/}, that represents polyhedra in a functionally similar way to
RVA, but much more concisely. The idea is to identify in the
transition relation of RVA structures that can be described
efficiently in algebraic notation, and to replace these structures by
their implicit representation. We show that checking whether a vector
belongs to a set represented by an IRVA can be decided very efficiently, by
following a single path in its transition graph.  We also develop
algorithms for minimizing an IRVA into a canonical form, and for
applying Boolean operators to IRVA.

\section{Basic Notions}

\subsection{Linear Constraints and Polyhedra}
\label{sec-constraints}

Let $n \in \nats_{>0}$ be a dimension.  A {\em linear constraint\/}
over vectors $\vec{x} \in \reals^n$ is a constraint of the form
$\vec{a}.\vec{x} \,\#\, b$, with $\vec{a} \in \ints^n$, $b \in \ints$,
and $\#\in\{<, \leq, =, \geq,$ $> \}$. A finite Boolean combination of
such constraints forms a {\em polyhedron\/}. If a polyhedron can be
expressed as a finite conjunction of linear constraints, it is said to
be {\em convex\/}.  A polyhedron that can be expressed as a
conjunction of linear equalities, i.e., constraints of the form
$\vec{a}.\vec{x} = b$, is an {\em affine space\/}. An affine space
that contains $\vec{0}$ is a {\em vector space\/}. The {\em
  dimension\/} $\mbox{dim}(\mbox{VS})$ of a vector space $\mbox{VS}$
is the size of the largest set of linearly independent vectors it
contains.

 Finally, given a convex polyhedron $D \subseteq \reals^n$, a
 polyhedron $P \subseteq \reals^n$, and a vector $\vec{v} \in D$, we
 say that $P$ is {\em conical\/} in $D$ with respect to the {\em
 apex\/} $\vec{v}$ iff for all $\vec{x} \in D$ and
 $\lambda\in]0,1[$, we have $\vec{x} \in P \,\Leftrightarrow\,
 \lambda(\vec{x} - \vec{v}) + \vec{v} \in P$. (Intuitively, this
condition expresses that within $D$, the polyhedron $P$ is not
affected by a scaling centered on $\vec{v}$.)  It is shown
in~\cite{BBL09} that the set of the vectors $\vec{v}$ with respect to
which $P$ is conical in $D$ necessarily coincides with an affine space
over $D$.

\subsection{Real Vector Automata}

This section is adapted from~\cite{BBR97,BJW05,BBL09}.  Let $r \in
\nats_{>1}$ be a {\em numeration base\/}. In the positional number
system in base $r$, a number $z \in \reals_{\geq 0}$ can be {\em
  encoded\/} by an infinite word $a_{p-1} a_{p-2} \ldots a_0 \star
a_{-1} a_{-2} a_{-3} \ldots$, where $\forall i:\, a_i \in \{ 0, 1,
\ldots, r-1 \}$, such that $z = \sum_{i < p} a_i r^i$. (The
distinguished symbol ``$\star$'' separates the integer from the
fractional part of the encoding.) Negative numbers are encoded by
using the {\em $r$'s-complement\/} method, which amounts to
representing a number $z \in \reals_{< 0}$ by the encoding of $z +
r^p$, where $p$ is the length of its integer part. This length $p$
does not have to be fixed, but must be large enough for the constraint
$-r^{p-1} \leq z \leq r^{p-1}$ to hold, in order to reliably
discriminate the sign of encoded numbers. Under this scheme, every
real number admits an infinite number of encodings in base $r$. Note
that some numbers admit different encodings with the same integer-part
length, for instance, the base-$2$ encodings of $1/4$ form the
language $0^+\star 0 1 0^{\omega}\,\cup\, 0^+ \star 0 0 1^{\omega}$.
Such encodings are then called {\em dual\/}.

The positional encoding of numbers generalizes to vectors in
$\reals^n$, with $n \in \nats_{>0}$. A vector is encoded by first
choosing encodings of its components that share the same integer-part
length. Then, these component encodings are combined by repeatedly and
synchronously reading one symbol in each component. The result takes
the form of an infinite word over the alphabet $\{ 0, 1, \ldots, r-1
\}^n \,\cup\, \{ \star \}$ (since the separator is read simultaneously
in all components, it can be denoted by a single symbol). It is also
worth mentioning that the exponential size of the alphabet can be
avoided if needed by {\em serializing\/} the symbols, i.e., reading the
components of each symbol sequentially in a fixed order rather than
simultaneously~\cite{Boi98}.

This encoding scheme maps any set $S \subseteq \reals^n$ onto a language of
infinite words. If this language is $\omega$-regular, then it can be
accepted by an infinite-word automaton, which is then known as a
{\em Real Vector Automaton (RVA)\/} representing the set $S$.

Some classes of infinite-word automata are notoriously difficult to handle
algorithmically~\cite{Var07}. A {\em weak\/} automaton is a Büchi automaton such
that each strongly connected component of its transition graph
contains either only accepting or only non-accepting states. The
advantage of this restriction is that weak automata admit efficient
manipulation algorithms, comparable in cost to those suited for
finite-word automata~\cite{Wil93}. The following result is established
in~\cite{BJW05}.

\begin{theorem}
\label{theo-repr}
Let $n \in \nats_{>0}$. Every polyhedron of $\reals^n$ can be
represented by a weak deterministic RVA, in every base $r \in
\nats_{>1}$.
\end{theorem}

In the sequel, we will only consider weak and deterministic RVA. These
structures can efficiently be minimized into a canonical
form~\cite{Lod01}, and combining them by Boolean operators amounts to
performing similar operations on the languages they
accept. Implementations of RVA are available as parts of the tools
LASH~\cite{LASH} and LIRA~\cite{LIRA}.

\section{The Structure of Polyhedra}
\label{sec-stucture-polyhedra}

It is known that RVA can form unnecessarily large representations of
polyhedra. For instance, a finite-state automaton recognizing the set
of solutions $(x_1, x_2)$ of the constraint $x_1 = r^k x_2$ in base
$r$ essentially has to check that $x_1$ and $x_2$ have identical
encodings up to a shift by $k$ symbols, and thus needs $O(r^k)$ states
for its memory. On the other hand, the algebraic description of the 
constraint $x_1 = r^k x_2$ requires only $O(k)$ symbols.

In this section, we study the transition relation of RVA representing
polyhedra, with the aim of finding internal structures that can more
efficiently be described in algebraic notation.

\subsection{Conical Sets}
\label{sec-conical-sets}

It has been observed in~\cite{GHH+03} that, for every
polyhedron $P \subseteq \reals^n$ and point $\vec{v} \in \reals^n$,
the set $P$ is conical in all sufficiently small convex neighborhoods
of $\vec{v}$.  We now formalize this property, and prove it by
reasoning about the structure of RVA representing $P$. This will
provide valuable insight into the principles of operation of
automata-based representations of polyhedra.

For every $\vec{v} = (v_1, v_2, \ldots, v_n)\in\reals^n$ and
$\varepsilon\in\reals_{>0}$, let
$N_{\varepsilon}(\vec{v})$ denote the $n$-cube of size $[\varepsilon]^n$
centered on $\vec{v}$, that is, the set
$[v_1 - \varepsilon/2, v_1 +
  \varepsilon/2] \times [v_2 - \varepsilon/2, v_2 + \varepsilon/2]
\times \cdots \times [v_n - \varepsilon/2, v_n + \varepsilon/2]$.

\begin{theorem}
\label{theo-conical}
Let $P \subseteq \reals^n$ be a polyhedron, with $n \in \nats_{>0}$,
and let $\vec{v}\in\reals^n$ be an arbitrary point. For every
sufficiently small $\varepsilon \in \reals_{>0}$, the set $P$ is
conical in $N_{\varepsilon}(\vec{v})$ with respect to the apex
$\vec{v}$.
\end{theorem}
\proof\ Let $\cal A$ be a RVA representing $P$ in a base $r \in
\nats_{>1}$, which exists thanks to Theorem~\ref{theo-repr}.  We
assume w.l.o.g. that $\cal A$ is weak, deterministic, and has a
complete transition relation. Consider a word $w$ encoding $\vec{v}$
in base $r$. For each $k \in \nats$, let $w_k$ denote the finite
prefix of $w$ with $k$ symbols in its fractional part, i.e., such that
$w_k = u \star u'$ with $|u'| = k$.  The set of all vectors that admit
an encoding of prefix $w_k$ forms a $n$-cube $C_{w_k}$ of size
$[r^{-k}]^n$. For every $k \in \nats$, we have $\vec{v} \in C_{w_k}$
and $C_{w_{k+1}} \subset C_{w_k}$, leading to $\bigcap_{k\in\nats}
C_{w_k} = \{ \vec{v} \}$.  Intuitively, each symbol read by $\cal A$
reduces by a factor $r^n$ the size of the set of possibly recognized
vectors.

Consider $\varepsilon \in \reals_{>0}$ with $\varepsilon < 1$. The set
$N_{\varepsilon}(\vec{v})$ is covered by the union of the sets
$C_{w_k}$ for all $w$ encoding $\vec{v}$, choosing $k$ such that
$r^{-k} \geq \varepsilon $. It is thus sufficient to prove that for
every word $w$ encoding $\vec{v}$ and sufficiently large $k$, the set
$P$ is conical in $C_{w_k}$ with respect to the apex $\vec{v}$.  This
property has been proved in~\cite{BBL09}, where it is additionally
shown that the suitable values of $k$ include those for which $w_k$
reaches the last strongly connected component of $\cal A$ visited by
$w$.  \qed

In the previous proof, the strongly connected components of $\cal A$
turn out to be connected to conical structures present in $P$. This
can be explained as follows. Consider two finite prefixes $w_k$ and
$w_{k+d}$ of $w$, with $d > 0$, such that $w_{k+d}$ only differs from
$w_k$ by additional iterations of cycles in the last strongly
connected component of $\cal A$ visited by $w$.  Since both $w_k$ and
$w_{k+d}$ lead to the same state of $\cal A$, the sets of suffixes that can be
appended to them so as to obtain words accepted by $\cal A$ are
identical. In order to be able to compare such sets of suffixes,
we introduce the following notation. For each $k \in \nats$, let $\vec{c}_{w_k} = (c_{w_{k},1}, c_{w_{k},2}, \ldots, c_{w_{k}, n})$ denote the vector encoded
by $w_k (0^n)^{\omega}$, in other words the vector such that
$C_{w_k} = [c_{w_{k},1}, c_{w_{k},1} + r^{-k}] \times [c_{w_{k},2}, c_{w_{k},2}
+ r^{-k}] \times \cdots \times [c_{w_{k},n}, c_{w_{k},n} + r^{-k}]$. 
Given a $n$-cube $C \subset \reals^n$ of size $[\lambda]^n$ and a
vector $\vec{c} \in \reals^n$, we then define the
{\em normalized view\/} of $P$ with respect to $C$ and $\vec{c}$ as the
set $P[C, \vec{c}] = (1/\lambda)((P \,\cap\, C) - \vec{c})$.
In other words, this normalized view is
obtained by a translation bringing $\vec{c}$ onto the origin
$\vec{0}$, followed by a scaling that makes the size of the $n$-cube
in which $P$ is observed become equal to $[1]^n$.

Observe that the set $P[C_{w_k}, \vec{c}_{w_k}]$ is precisely
characterized by the language accepted from the state of $\cal A$
reached by $w_k$.  Since this state is identical to the one reached by
$w_{k+d}$, we obtain $P[C_{w_k}, \vec{c}_{w_k}] = P[C_{w_{k+d}},
  \vec{c}_{w_{k+d}}]$.  Recall that we have $\vec{v} \in C_{w_k}$ and
$C_{w_{k+d}} \subset C_{w_k}$. The previous property shows that $P$ is
{\em self-similar\/} in the vicinity of $\vec{v}$: Following
additional cycles in the last strongly connected component visited by
$w$ amounts to increasing the ``zoom level'' at which the set $P$ is
viewed close to $\vec{v}$, without influencing this view. It is shown
in~\cite{BBL09} that this self-similarity entails the conical
structure of $P$ around $\vec{v}$, which intuitively means that the
zoom levels that preserve the local structure of $P$ are not
restricted to integer powers of $r^d$.

In addition, we have established that the structure of $P$ in a small
neighborhood $N_{\varepsilon}(\vec{v})$ of $\vec{v}$ is uniquely
determined by the state of $\cal A$ reached by $w_k$. Since there are
only finitely many such states, we have the following result.

\begin{theorem}
\label{theo-local-view}
Let $P \subseteq \reals^n$ be a polyhedron, with $n \in \nats_{>0}$.
There exists $\varepsilon \in \reals_{>0}$ such that over all points
$\vec{v}\in\reals^n$, the sets $P
[N_{\varepsilon}(\vec{v}), \vec{v}]$ take a finite number
of different values. Moreover, each of these sets is conical in
$[-1/2, 1/2]^n$ with respect to the apex $\vec{0}$.
\end{theorem}
\proof\ The proof follows the same lines as the one of
Theorem~\ref{theo-conical}. Let $\cal A$ be a weak, deterministic,
and complete RVA representing $P$ in a base $r \in \nats_{>1}$.  To
every word $w$ encoding a given vector $\vec{v}$ in base $r$, we
associate the integer $k(w)$ such that the path of $\cal A$
recognizing $w$ reads the finite prefix $w_{k(w)}$ before reaching the
last strongly connected component that it visits. From the previous
developments, we have that $P$ is conical in $C_{w_{k(w)}}$ with
respect to the apex $\vec{v}$. Furthermore, the set $P[C_{w_{k(w)}},
  \vec{c}_{w_{k(w)}}]$ only depends on the state of $\cal A$ reached
after reading $w_{k(w)}$, which are in finite number. It follows that,
in arbitrarily small neighborhoods of $\vec{v}$, the polyhedron $P$
has a conical structure with respect to the apex $\vec{v}$, and that
there are only finitely many such structures over all vectors $\vec{v}$.
\qed

\subsection{Polyhedral Components}
\label{sec-components}

Theorem~\ref{theo-local-view} shows that a polyhedron $P \subseteq
\reals^n$ partitions $\reals^n$ into finitely many equivalence
classes, each of which corresponds to a unique conical set in the
$n$-cube $[-1/2, 1/2]^n$ with respect to the apex $\vec{0}$. For each
$\vec{v} \in \reals^n$, let $P_{\vec{v}} \subseteq [-1/2, 1/2]^n$ denote
the conical set associated to $\vec{v}$ by $P$. We call $P_{\vec{v}}$
the {\em component\/} of $P$ associated to $\vec{v}$. Recall that,
as discussed in Section~\ref{sec-constraints}, the
set of apexes according to which $P_{\vec{v}}$ is conical coincides
with a vector space over $[-1/2, 1/2]^n$. The {\em dimension\/}
$\mbox{dim}(P_{\vec{v}})$ of the component $P_{\vec{v}}$ is defined as
the dimension of this vector space. Finally, we say that a component
$P_{\vec{v}}$ is {\it in\/} if $\vec{v} \in P$, and {\it out\/}
if $\vec{v} \not\in P$.

\begin{figure}
\centerline{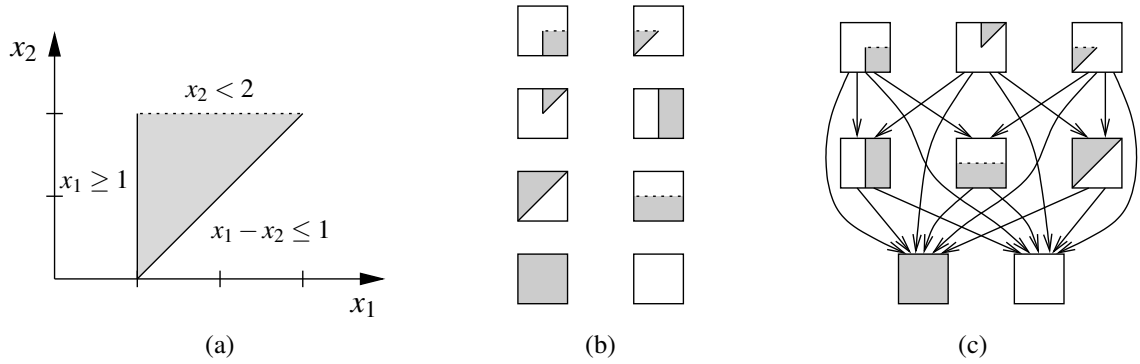}
\caption{Example of (a) polyhedron, (b) polyhedral components, and 
(c) incidence relation.}
\label{fig-example-components}
\end{figure}

An example is given in Figure~\ref{fig-example-components}. The
triangle $x_1 \geq 1 \,\wedge\, x_2 < 2 \,\wedge\, x_1 - x_2 \leq 1$
in $\reals^2$ has three components of dimension $0$ corresponding to
its vertices $(1, 0)$ ({\it in\/}), $(1, 2)$ ({\it out\/}) and $(3,
2)$ ({\it out\/}), three components of dimension $1$ associated to its
sides (two {\it in\/} and one {\it out\/}), and two components of
dimension $2$ corresponding to its interior ({\it in\/}) and exterior
({\it out\/}) points.

\subsection{Incidence Relation}

In Section~\ref{sec-conical-sets}, we have established a link between
the components of a polyhedron $P \subseteq \reals^n$ and the strongly
connected components (SCC) of a RVA $\cal A$ representing $P$. We know
that there exists a hierarchy between the SCC of an automaton: That a
SCC $S_2$ is reachable from a SCC $S_1$ implies that every finite
prefix that reaches a state of $S_1$ can be followed by a suffix that
ends up visiting $S_2$, while the reciprocal property does not
hold. In a similar way, we can define an {\em incidence relation\/}
between the components of a polyhedron.

\begin{definition}
Let $Q_1, Q_2$ be distinct components of a polyhedron $P \subset
\reals^n$, with $n \in \nats_{>0}$. The component $Q_2$ is {\em
  incident\/} to $Q_1$, denoted $Q_1 \prec Q_2$, iff for all
$\vec{v}_1 \in \reals^n$ such that $P_{\vec{v}_1} = Q_1$ and
$\varepsilon \in \reals_{>0}$, there exists $\vec{v}_2 \in \reals^n$
such that $P_{\vec{v}_2} = Q_2$ and $|\vec{v}_1 - \vec{v}_2| <
\varepsilon$.
\end{definition}

Remark that the incidence relation between the components of a
polyhedron is a partial order, and that $Q_1 \prec Q_2$ implies
$\mbox{dim}(Q_1) < \mbox{dim}(Q_2)$. As an example, in the triangle
depicted in Figure~\ref{fig-example-components}, each side is
incident to the vertices it links, since every neighborhood of a
vertex contains points from its adjacent sides. The reverse property
does not hold.  The interior and exterior components of the triangle
are incident to each of its sides and vertices.

\subsection{How RVA Recognize Vectors}
\label{sec-rva-decision}

We are now able to explain the mechanism employed by a RVA $\cal A$ in
order to check whether the vector encoded by a word $w$ belongs or not
to a polyhedron $P \subseteq \reals^n$. After reading an integer part
and a separator symbol, the word $w$ follows some transitions in the
fractional part of $\cal A$, reaching a first non-trivial strongly
connected component $S_1$ (that is, a component containing at least
one cycle). At this location in $w$, inserting arbitrary iterations of
cycles within $S_1$ would not affect the accepting status of $w$. This
intuitively means that the prefix $w_k$ of $w$ read so far has led us
to a point that belongs to a component $Q_1$ of $P$, and that the
decision can now be carried out further in an arbitrarily small
neighborhood of this point. Reading additional symbols from $w$, one
either stays within $S_1$, or follows transitions that eventually lead
to another non-trivial strongly connected component $S_2$. Once again,
this means that the decision can now take place in an arbitrarily
small neighborhood of a point belonging to a component $Q_2$ of $P$,
such that either $Q_1 = Q_2$ or $Q_1 \prec Q_2$. The same procedure
repeats itself until $w$ reaches a strongly connected component that
it does not leave anymore.

In other words, in order to decide whether to accept or not a word
$w$, the RVA $\cal A$ first chooses deterministically a component
$Q_1$ of $P$ in the vicinity of which this decision can be carried
out. Then, it checks whether the vector $\vec{v}$ encoded by $w$
belongs or not to $Q_1$. If yes, the decision is taken according to
whether $Q_1$ is {\it in\/} or {\it out\/}. If no, the RVA chooses
deterministically a component $Q_2$ incident to $Q_1$, from which the
same procedure is then repeated.

Let us now study more finely the mechanism used for moving from a
component $Q_1$ that does not contain the vector $\vec{v}$ to another
component $Q_2$ from which $\vec{v}\in P$ can be decided. One follows
a path of $\cal A$ that leaves a strongly connected component
associated to $Q_1$, travels through an acyclic structure of
transitions, and finally reaches a SCC associated to $Q_2$. Recall
that, as discussed in Section~\ref{sec-conical-sets}, at each step in
this path, the prefix $w_k$ of $w$ read so far determines a $n$-cube
$C_{w_k}$. This $n$-cube covers some subset $U_{w_k} = \{ P_{\vec{u}}
\mid \vec{u} \in C_{w_k}\}$ of the components of $P$. If $U_{w_k}$
contains a single minimal component with respect to the incidence
order $\prec$, then this component is necessarily equal to $Q_2$, and
its associated SCC is the only possible destination of $w_k$.
Indeed, all components in $U_{w_k}$ are then either equal or incident to
$Q_2$. If, on the other hand, $U_{w_k}$ contains more than one minimal
component, then further transitions have to be followed in order to
discriminate between them.

\section{Implicit Real Vector Automata}

Our goal is to define a data structure representing a polyhedron $P
\subseteq \reals^n$ that is more concise than a RVA, but from which
one can decide $\vec{v}\in P$ using a similar procedure to the one
outlined in Section~\ref{sec-rva-decision}.  There are essentially
three operations to consider: Selecting from a vector $\vec{v}$ an
initial polyhedral component from which the decision can be started,
checking whether $\vec{v}$ belongs or not to a given component, and
moving from a component that does not contain $\vec{v}$ to another one
from which the decision can be continued. We study separately each of
these problems in the three following sections.

\subsection{Choosing an Initial Component}
\label{sec-initial-component}

An easy way of managing the choice of an initial component is to
consider only polyhedra in which this component is unique. This can be
done without loss of generality thanks to the following definition.

\begin{definition}
Let $P \subseteq \reals^n$ be a polyhedron, with $n \in \nats_{>0}$.
The {\em representing cone\/} of $P$ is the polyhedron 
$\overline{P} \subseteq \reals^{n+1} = \{ \lambda (x_1,
\ldots, x_n, 1)
 \mid \lambda \in \reals_{>0} \,\wedge\, (x_1, \ldots, x_n) \in P\}$.
\end{definition}

For every polyhedron $P \subseteq \reals^n$, the polyhedron
$\overline{P}$ is conical in $\reals^{n+1}$ with respect to the apex
$\vec{0}$, from which it can be inferred that every neighborhood of
$\vec{0}$ contains a unique minimal component $Q_0$ with respect to the
incidence order $\prec$. It follows that for every $\vec{v}\in
\reals^{n+1}$, the decision $\vec{v} \in \overline{P}$ can be started
from $Q_0$. Remark that $\overline{P}$ describes $P$ without
ambiguity, since $P$ can be reconstructed from $\overline{P}$ by
computing its intersection with the constraint $x_{n+1} = 1$, and
projecting the result over the first $n$ vector components.  In the
sequel, we assume w.l.o.g. that the polyhedra that we consider are
conical with respect to the apex $\vec{0}$. A similar mechanism is employed
in~\cite{LeV92}.

\subsection{Deciding Membership in a Component}
\label{sec-decision-inside}

Consider a polyhedron $P \subseteq \reals^n$ that is conical with
respect to the apex $\vec{0}$. As explained in
Section~\ref{sec-components}, a component of such a polyhedron is
characterized by a vector space, a Boolean polarity (either {\it in\/}
or {\it out\/}), and its incident components. Checking whether a given
vector $\vec{v} \in \reals^n$ belongs or not to the component reduces
to deciding whether $\vec{v}$ belongs to its associated vector
space. This is a simple algebraic operation if, for instance, the
vector space is represented by a vector basis $\{\vec{b}_1, \vec{b}_2,
\ldots, \vec{b}_m \}$: One simply has to check whether $\vec{v}$ is
linearly dependent with $\{\vec{b}_1, \vec{b}_2, \ldots, \vec{b}_m \}$.
 This approach leads to a much more concise representation
of polyhedral components than the one used in RVA.

\subsection{Moving from a Component to Another}
\label{sec-decision-moving}

We now address the problem of leaving a component $Q_1$ of a
polyhedron $P\subseteq\reals^n$ that does not contain a vector
$\vec{v}\in\reals^n$, and moving to a component $Q_2$ that is incident
to $Q_1$, and from which $\vec{v} \in P$ can be decided.

A first solution would be to borrow from a RVA representing $P$ the
acyclic structure of transitions leaving the strongly connected
components $S_1$ associated to $Q_1$. However, this would negate the
advantage in conciseness obtained in
Section~\ref{sec-decision-inside}, since this acyclic structure of
transitions is generally as large as $S_1$ itself.

The solution we propose consists in performing a variable change
operation. Let $\{\vec{y}_1, \vec{y}_2, \ldots, \vec{y}_m\}$, with $0
< m \leq n$, be a basis of the vector space associated with the
component $Q_1$. If $m = n$, then $Q_1$ is universal and there is no
possibility of leaving it. If $m < n$, then we introduce $n-m$ additional
vectors $\vec{z}_1$, $\vec{z}_2$, \ldots, $\vec{z}_{n-m}$, such that
$\{ \vec{y}_1, \ldots, \vec{y}_m, \vec{z}_1, \ldots, \vec{z}_{n-m} \}$
forms a basis of $\reals^n$. These additional vectors can be chosen in
a canonical way by selecting among $(1,0, \ldots, 0), (0,
1, \ldots, 0), \ldots, (0, 0, \ldots, 1)$, considered in that order,
$n-m$ vectors that are linearly independent with $\{ \vec{y}_1, \vec{y}_2,
\ldots, \vec{y}_m \}$.

We then express the vector $\vec{v}$ in the coordinate system $\{
\vec{y}_1, \ldots, \vec{y}_m, \vec{z}_1, \ldots,$ $\vec{z}_{n-m} \}$,
obtaining a vector $(y_1, \ldots, y_m, z_{1}, \ldots, z_{n-m})$. That
$\vec{v}$ leaves $Q_1$ simply means that we have $(z_1, \ldots,
z_{n-m}) \neq \vec{0}$.  As a consequence, we associate $Q_1$ with an
acyclic structure ${\cal D}_1$ of outgoing transitions, recognizing
prefixes of encodings of non-zero vectors $(z_{1}, \ldots, z_{n-m})$,
in order to map these vectors to the polyhedral components (incident to
$Q_1$) to which they lead.

A difficulty is that, from Theorem~\ref{theo-conical}, the set $P$ has
a conical structure in arbitrary small neighborhoods of points in
$Q_1$. If follows that the structure ${\cal D}_1$ has to map onto the
same polyhedral component two vectors $\vec{z}$ and $\vec{z}'$ such
that $\vec{z}' = \lambda \vec{z}$ for some $\lambda \in \reals_{>0}$.
An efficient solution is to {\em normalize\/} the vectors handled by
${\cal D}_1$: Given a vector $\vec{z} = (z_{1}, \ldots, z_{n-m})$ such
that $\vec{z} \neq \vec{0}$, we define its normalized form as
$[\vec{z}] = (1/{(2.\mbox{max}_i |z_i|)}) \vec{z}$.  In other words,
$[\vec{z}]$ is obtained by turning $\vec{z}$ into the half-line $\{
\lambda \vec{z} \mid \lambda \in \reals_{>0}\}$, and computing the
intersection of this half-line with the faces of the {\em
  normalization cube\/} $[ -1/2, 1/2 ]^{n-m}$. In this way, two
vectors that only differ by a positive factor share the same
normalized form, and will thus be handled identically.

The purpose of the structure ${\cal D}_1$ is thus to recognize
normalized forms of vectors, and map them onto the polyhedral
components to which they lead. In order to define the transition graph
of ${\cal D}_1$, one therefore needs a suitable encoding for
normalized forms of vectors. Using the standard positional encoding of
vectors in a base $r \in \nats_{>1}$ is possible, but inefficient.  We
instead use the following scheme. An encoding of a normalized vector
$[\vec{v}] = ([v]_1, [v]_2 \,\ldots, [v]_{n-m})$ starts with a leading
symbol $a \in \{ -1, +1, -2, +2, \ldots, -(n-m), +(n-m) \}$ that
identifies the face of the normalization cube $[ -1/2, 1/2 ]^{n-m}$ to
which $[\vec{v}]$ belongs: If $a = -i$, with $1 \leq i \leq n-m$, then
$[v]_i = -1/2$; if $a = +i$, then $[v]_i = +1/2$. This prefix is
followed by a suffix $w \in \{ 0, 1 \}^{\omega}$ that encodes the
position of $[\vec{v}]$ within the face of the normalization cube
defined by $a$. This suffix is obtained as follows. Assume that we
have $a \in \{ -i, +i \}$, with $1 \leq i \leq n-m$ (which implies
$[v]_i \in \{ -1/2, 1/2 \}$).  We turn $[\vec{v}]$ into $[[\vec{v}]] =
([v_1], \ldots, [v_{i-1}], [v_{i+1}], \ldots, [v_{n-m}]) + (1/2, 1/2,
\ldots, 1/2)$, i.e., we remove the $i$-th vector component, and offset
the result in order to obtain $[[\vec{v}]] \in [0, 1]^{n-m-1}$. We then
define $w \in \{ 0, 1 \}^{\omega}$ as a word such that $0 \star w$ is
a serialized binary encoding of $[[\vec{v}]]$. Note that some vectors
$\vec{v}$ may belong to several faces of the normalization cube, hence
their normalized form may admit multiple encodings. This is not
problematic, provided that the structure ${\cal D}_1$ handles these
encodings consistently.

In summary, the structure ${\cal D}_1$ is an acyclic decision graph
that partitions the space of normalized vectors according to their
destination components. Each prefix $w_k$ of length $k$ read by ${\cal
  D}_1$ corresponds to a convex region $R_{w_k} \subset \reals^n$ that
is conical in every neighborhood of any element of $Q_1$, with this
element as apex.  The situation is similar to that discussed in
Section~\ref{sec-rva-decision}: If in a sufficiently small
neighborhood of any point of $Q_1$, the set of components of $P$
covered by $R_{w_k}$ contains a unique minimal component $Q_2$ with
respect to the incidence order $\prec$, then $w_k$ leads to
$Q_2$. Otherwise, the decision process is not yet complete, and
additional transitions have to be followed in ${\cal D}_1$.

\subsection{Data Structure}

We are now ready to describe our proposed data structure for representing
arbitrary polyhedra of $\reals^n$, with $n \in \nats_{>0}$. Recall that
we assume w.l.o.g. that the polyhedra we consider are conical in
$\reals^n$ with respect to the apex $\vec{0}$.

\subsubsection{Syntax}

\begin{definition}
An\/ {\em Implicit Real Vector Automaton (IRVA)} is a tuple 
$(n, S_I, S_E,$ $s_0, \Delta)$, where
\begin{itemize}
\item $n$ is a\/ {\em dimension}.
\item $S_I$ is a set of\/ {\em implicit states}. Each $s \in S_I$ is
associated with a\/ {\em vector space} $\mbox{VS}(s) \subseteq \reals^n$,
and a Boolean\/ {\em polarity} $\mbox{pol}(s) \in \{ \mbox{\it in\/},
\mbox{\it out\/} \}$.
\item $S_E$ is a set of\/ {\em explicit states}, such that $S_E
\,\cap\, S_I = \emptyset$.
\item $s_0 \in S_I$ is the\/ {\em initial state}.
\item $\Delta :\,S_I \times \pm \nats_{>0} ~\cup~
S_E \times \{ 0, 1 \} \rightarrow
(S_I \,\cup\, S_E)$ is a (partial)\/ {\em transition relation}.
\end{itemize}
\end{definition}

In order to be well formed, an IRVA $(n, S_I, S_E, s_0, \Delta)$
representing a polyhedron $P \subseteq \reals^n$ has to satisfy some
integrity constraints. In particular, the transition relation $\Delta$
must be acyclic, and for all $s_1, s_2 \in S_I$ such that $\Delta$
directly or transitively leads from $s_1$ to $s_2$, one must have
$\mbox{VS}(s_1) \subset \mbox{VS}(s_2)$. The transition relation
$\Delta$ is required to be complete, in the sense that, for every
implicit state $s \in S_I$ and $i \in \nats_{>0}$, $\Delta(s, +i)$ and
$\Delta(s, -i)$ are defined iff $i \leq n -
\mbox{dim}(\mbox{VS}(s))$. Furthermore, for every explicit state $s
\in S_E$, both $\Delta(s, 0)$ and $\Delta(s, 1)$ must be
defined. Finally, each component of $P$ must be described by a state
in $S_I$, and for every pair $Q_1, Q_2$ of components of $P$ such that
$Q_1 \prec Q_2$, there must exist a sequence of transitions in
$\Delta$ leading from the implicit state associated to $Q_1$ to the
one associated to $Q_2$. In other words, the order $\prec$ between the
components of $P$ can straightforwardly be recovered from the
reachability relation between the implicit states representing them.

\subsubsection{Semantics}
\label{sec-semantics}

The semantics of IRVA is defined by the following procedure, that
decides whether a given vector $\vec{v} \in \reals^n$ belongs or not
to the polyhedron $P$ represented by an IRVA $(n, S_I, S_E, s_0,
\Delta)$. The principles of this procedure have already been outlined
in Sections~\ref{sec-decision-inside} and~\ref{sec-decision-moving}.

One starts at the implicit state $s_0$. At each visited implicit state
$s$, one first decides whether $\vec{v} \in \mbox{VS}(s)$.  In case of
a positive answer, the procedure concludes that $\vec{v} \in P$ if
$\mbox{pol}(s) = \mbox{\it in\/}$, and that $\vec{v} \not\in P$
otherwise. In the negative case, the decision has to be carried out
further.  The vector $\vec{v}$ is transformed into $\vec{v}'$
according to the variable change operation associated to
$\mbox{VS}(s)$.  Then, $\vec{v}'$ is normalized into a vector
$[\vec{v}']$, which is encoded into a word $w \in \pm\nats \{ 0, 1
\}^{\omega}$. (In the case of multiple encodings, one of them can
arbitrarily be chosen.) The word $w$ corresponds to a single path of
transitions leaving $s$, which is followed until a new implicit state
$s'$ is reached. Note that the states visited by this path 
between $s$ and $s'$ are explicit ones. The procedure then repeats itself from 
this state $s'$.

\subsubsection{Examples}

\begin{figure}
\centerline{\small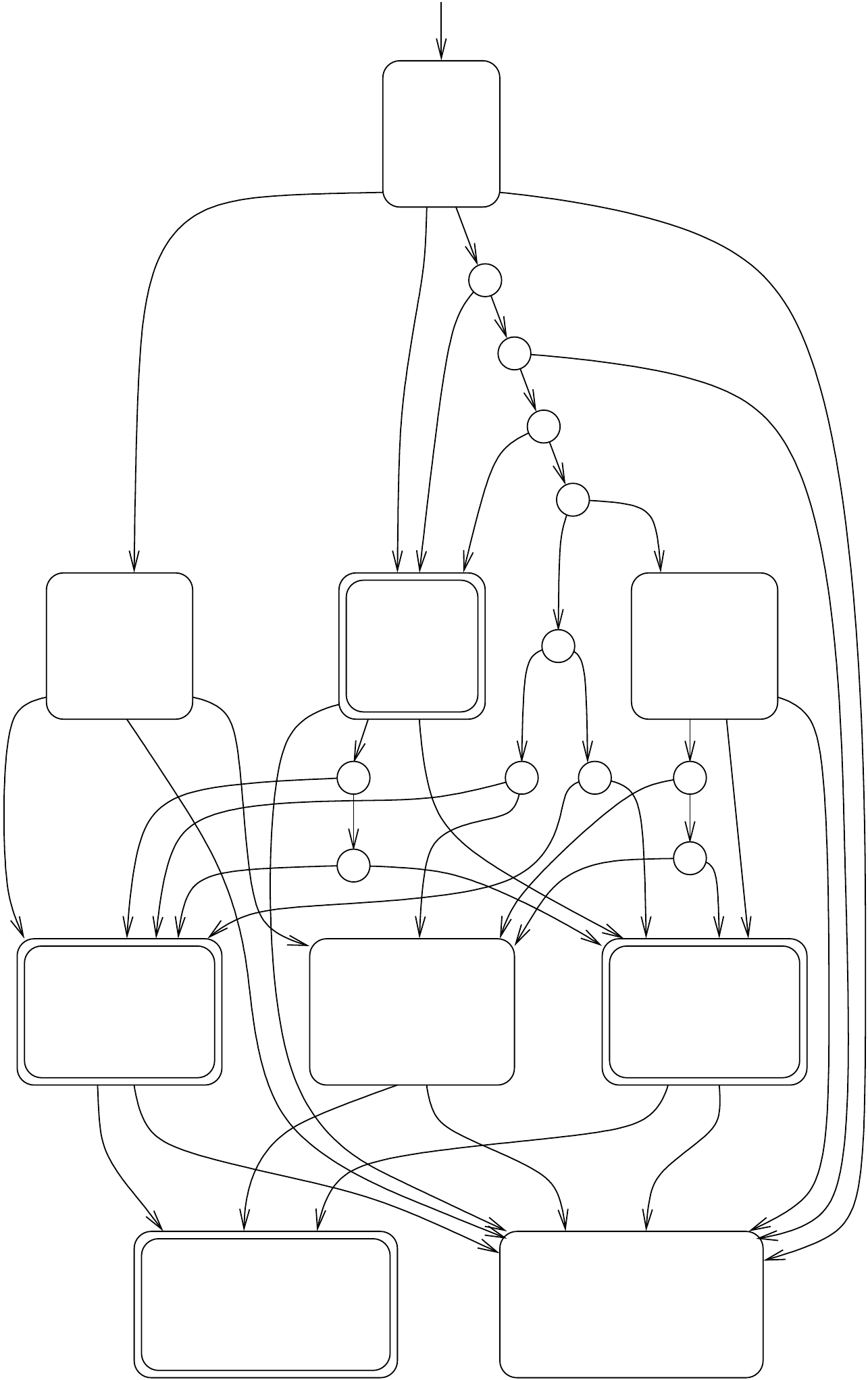}
\caption{IRVA representing the set  $\{(x_1,x_2) \in \reals^2 \mid 
x_1 \geq 1 \,\wedge\, x_2 < 2 \,\wedge\,
x_1 - x_2 \leq 1\}$.}
\label{fig-example-irva}
\end{figure}

An IRVA representing the set $x_1 \geq 1 \,\wedge\, x_2 < 2 \,\wedge\,
x_1 - x_2 \leq 1$ in $\reals^2$, considered in
Figure~\ref{fig-example-components}(a), is given in
Figure~\ref{fig-example-irva}.  Note that, since the set is not
conical, the IRVA actually recognizes its representing cone, as
discussed in Section~\ref{sec-initial-component}. In this figure,
implicit states are depicted by rounded boxes, and explicit ones by
small circles. Doubled boxes represent {\em in\/} polarities. The
vector spaces associated to implicit states are represented by one of
their bases. Remark that the layout of the implicit states and the
decision structures linking them closely matches the polyhedral
components and their incidence relation as depicted in
Figure~\ref{fig-example-components}(c), except for the initial state
which corresponds to the apex $\vec{0}$ of the representing cone.

\newcommand{\retrait}{\!\!\!}
\newcommand{\IRVAEVO}{ \small $
\left\{ \retrait
\begin{array}{c c c}
~ \\ ~ \\ ~ \\
\end{array}
\retrait
\right\}
$}

\newcommand{\EVI}[3]{\small$
~\left\{ \retrait \retrait
\begin{array}{c c c}
\left( \retrait \begin{array}{c} #1 \\ #2 \\ #3 \end{array} 
\retrait \right)
\end{array}
\retrait \retrait
\right\}
$}

\newcommand{\EVII}[6]{\small$
\left\{ \retrait \retrait
\begin{array}{c c c}
\left( \retrait \begin{array}{c} #1 \\ #2 \\ #3 \end{array} 
\retrait \right), \retrait \retrait \retrait&
\left( \retrait \begin{array}{c} #4 \\ #5 \\ #6 \end{array} 
\retrait \right)
\end{array}
\retrait \retrait
\right\}
$}

\newcommand{\IRVAEVA}{ \EVI{1}{2}{1} }
\newcommand{\IRVAEVB}{ \EVI{1}{0}{1} }
\newcommand{\IRVAEVC}{ \EVI{1}{2/3}{1/3} }
\newcommand{\IRVAEVAB}{ \EVII{1}{0}{1} {0}{1}{0} }
\newcommand{\IRVAEVAC}{ \EVII{1}{0}{0} {0}{1}{1/2} }
\newcommand{\IRVAEVBC}{ \EVII{1}{0}{1} {0}{1}{-1} }

\newcommand{\IRVAEVU}{\small$
\left\{ \retrait \retrait
\begin{array}{c c c}
\left( \retrait \begin{array}{c} 1 \\ 0 \\ 0 \end{array} 
\retrait \right), \retrait \retrait \retrait&
\left( \retrait \begin{array}{c} 0 \\ 1 \\ 0 \end{array}
\retrait \right), \retrait \retrait \retrait&
\left( \retrait \begin{array}{c} 0 \\ 0 \\ 1 \end{array}  
\retrait \right)
\end{array}
\retrait \retrait
\right\}
$}

\begin{figure}
\centerline{\small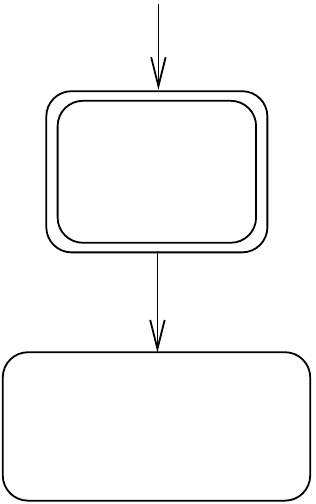}
\caption{IRVA representing the set $\{(x_1,x_2) \in \reals^2 \mid 
x_1 = 2^k x_2\}$.}
\label{fig-example-irva-rk}
\end{figure}

As an additional example, Figure~\ref{fig-example-irva-rk} shows how the 
set $x_1 = 2^k x_2$ in 
$\reals^2$, discussed in the 
introduction of Section~\ref{sec-stucture-polyhedra}, is represented by 
an IRVA.  In this case, the gain in conciseness is exponential with
respect to RVA.

\newcommand{\EVTI}[2]{\small$
~\left\{ \retrait \retrait
\begin{array}{c c c}
\left( \retrait \begin{array}{c} #1 \\ #2 \end{array} 
\retrait \right)
\end{array}
\retrait \retrait
\right\}
$}

\newcommand{\EVTII}[4]{\small$
\left\{ \retrait \retrait
\begin{array}{c c c}
\left( \retrait \begin{array}{c} #1 \\ #2 \end{array} 
\retrait \right), \retrait \retrait \retrait&
\left( \retrait \begin{array}{c} #3 \\ #4 \end{array} 
\retrait \right)
\end{array}
\retrait \retrait
\right\}$}

\newcommand{\IRVAEVRK}{ \EVTI{1}{1/2^k} }
\newcommand{\IRVAEVUU}{ \EVTII{1}{0} {0}{1} }

\section{Manipulation Algorithms}
\label{sec-algorithms}

\subsection{Test of Membership}

A procedure for checking whether a given vector belongs to a
polyhedron represented by an IRVA has already been outlined in
Section~\ref{sec-semantics}. In the case of a polyhedron $P \subseteq
\reals^n$ that is not conical, an IRVA can be obtained for its
representing cone $\overline{P} \subset \reals^{n+1}$, as discussed in
Section~\ref{sec-initial-component}. In this case, checking whether a
vector $(v_1, v_2, \ldots, v_n) \in \reals^n$ belongs to $P$ simply
reduces to determining whether $(v_1, v_2, \ldots, v_n, 1)$ belongs
to $\overline{P}$, which is done by the algorithm of
Section~\ref{sec-semantics}.

\subsection{Minimization}

An IRVA $(n, S_I, S_E, s_0, \Delta)$ can be minimized in order to
reduce its number of implicit and explicit states. Since the
transition relation $\Delta$ is acyclic, the explicit and implicit
states can be processed in a bottom-up order, starting from the
implicit states with the largest vector spaces. At each step,
reduction rules are applied in order simplify the current structure. A
first rule is aimed at merging states that are indistinguishable: If
two explicit states share the same successors, they can be merged. In
the case of two implicit states, one additionally has to check that
their associated vector spaces are equal, and that their polarities
match.  The purpose of the second rule is to get rid of unnecessary
decisions. Consider a state $s$ (either implicit or explicit) with an
outgoing transition that leads to an implicit state $s_1$,
representing a polyhedron component $Q_1$. If all the implicit states
$s_i$ that are reachable from $s$ are also reachable from $s_1$, then
these implicit states represent polyhedral components $Q_i$ such that
either $Q_i = Q_1$ or $Q_1 \prec Q_i$.  The state $s$ can then be
absorbed into $s_1$, provided that $s$ is not an implicit state with a
different polarity from the one of $s_1$. Note that this reduction
rule correctly handles the case of a state $s$ that is implicit and
does not correspond to a polyhedral component, but to a proper subset
of the component $Q_1$ represented by $s_1$. For example, in
$\reals^2$, $s$ may correspond to a unidimensional line $x_1 - x_2 =
0$ covered by the larger universal component $\reals^2$ represented by
$s_1$.

\begin{property}
Minimized IRVA are canonical up to isomorphism of their transition
relation, and equality of the vector spaces associated to their implicit
states.
\end{property}
\sketch\ The canonicity of a minimized IRVA ${\cal A}$ representing a
polyhedron $P \subseteq \reals^n$ is the consequence of two
properties. First, the minimization algorithm is able to identify and
merge together implicit states that correspond to identical polyhedral
components, as well as to remove the implicit states that do not
represent such components. This yields a one-to-one relationship
between the implicit states of $\cal A$ and the polyhedral components
of $P$.  Second, the transition structure leaving an explicit state
$s$ of $\cal A$ satisfies the following constraints. As discussed in
Section~\ref{sec-decision-moving}, the state $s$ corresponds to a
component $Q$ of $P$, and every prefix $w_k$ of length $k$ read from
$s$ defines a convex conical region $R_{s,w_k} \subset \reals^n$. If,
in all sufficiently small neighborhoods of $Q$, the region $R_{s,w_k}$
covers a unique component $Q'$ of $P$ that is minimal with respect to
the incidence order, then the path reading $w_k$ from $s$ leads to the
implicit state $s'$ corresponding to $Q'$. Provided that explicit
states that have identical successors are merged, this property
characterizes precisely the decision structure leaving $s$. Such
structures will then be isomorphic in all minimized IRVA representing
the same polyhedron.  \qed

\subsection{Boolean Combinations}

In order to apply a Boolean operator to two polyhedra $P_1$ and $P_2$
respectively represented by IRVA ${\cal A}_1$ and ${\cal A}_2$, one
builds an IRVA ${\cal A}$ that simulates the concurrent behavior of
${\cal A}_1$ and ${\cal A}_2$. The procedure is analogous to the
computation of the product of two finite-state automata. The initial
implicit state of ${\cal A}$ is obtained by combining the initial
states of ${\cal A}_1$ and ${\cal A}_2$, which amounts to intersecting
their associated vector spaces, and applying the appropriate Boolean
operator to their polarities. Each time an implicit state $s$ is added
to ${\cal A}$, representing a polyhedron component $Q$, its successors
are recursively explored. As explained in
Section~\ref{sec-decision-moving}, each finite prefix $w_k$, of length
$k$, read from $s$ corresponds to a convex conical region $R_{s,w_k}
\subset \reals^n$.  The idea is to check, in a sufficiently small
neighborhood $R$ of $Q$, whether $R_{s,w_k}$ covers unique minimal
components $Q_1$ of $P_1$ and $Q_2$ of $P_2$, with respect to their
respective incidence orders. In the positive case, one computes the
intersection of the underlying vector spaces of $Q_1$ and $Q_2$. If
the resulting vector space has a higher dimension than
$\mbox{dim}(\mbox{VS}(s))$, as well as a non-empty intersection with
$R_{s,w_k}$, a corresponding new implicit state is added to ${\cal
  A}$. In all other cases, the decision structure leaving $s$ has to
be further developed, which amounts to creating new explicit states and
new transitions between them, in order to read prefixes longer than
$w_k$.

A key operation in the previous procedure is thus to compute, from an
IRVA representing a polyhedron $P$, a component $Q$ of $P$, and a
given convex conical region $C$, the unique minimal component of $P$
(if it exists) covered by $C$ in the neighborhood of $Q$, with respect
to the incidence order $\prec$. This is done by exploring the IRVA
starting from the implicit state representing $Q$.  {From} a given
implicit state $s$, the exploration only has to consider the paths
labeled by words $w_k$ such that $C \,\cap\, R_{s,w_k} \neq
\emptyset$, until they reach another implicit state. Let $S$ be the
set of the implicit states reached this way. For each state in $S$,
one checks whether its underlying vector space has a non-empty
intersection with $C$. If this check succeeds for some nonempty subset
of $S$, then the procedure returns its minimal component, or fails
when such a component does not exist. Otherwise, it can be shown that
the exploration can be continued from a single state chosen
arbitrarily in $S$. The regions of space that are manipulated by this
procedure are convex polyhedra, and can be handled by specific data
structures~\cite{BRZH02}.

\section{Conclusions}

We have introduced a data structure, the Implicit Real Vector
Automaton (IRVA), that is expressive enough for representing arbitrary
polyhedra in $\reals^n$, closed under Boolean operators, and reducible
to a canonical form up to isomorphism.

IRVA share some similarities with the data structure described
in~\cite{GHH+03}, which also relies on decomposing polyhedra into
their components, and representing the incidence relation between
them. The main original feature of our work is the decision structures
that link each component to its incident ones, which are not limited to
three spatial dimensions, and lead to a canonical representation.
Furthermore, by imitating the
behavior of RVA, we have managed to obtain a symbolic representation
of polyhedra in which the membership of a vector can be decided by
following a single automaton path, which is substantially more
efficient that the procedure proposed in~\cite{GHH+03}.

The algorithms sketched in Section~\ref{sec-algorithms} are clearly
polynomial. We have not yet precisely studied their worst-case
complexity, since they depend on manipulations of convex polyhedra,
the practical cost of which is expected to be significantly lower than
their worst-case one. In order to assess the cost of building and
handling IRVA in actual applications, a prototype implementation of
those algorithms is under way. The example given in
Figure~\ref{fig-example-irva} has been produced by this prototype.

Future work will address other useful operations such as projection of
polyhedra, conversions to and from other representations, and
operations that are specific to symbolic state-space exploration
algorithms. For this particular application, IRVA in their present
form are still impractical, since they only provide efficient
representations of polyhedra in spaces of small dimension. (Indeed, the
size of an IRVA grows with the number of components of the polyhedron
it represents, and simple polyhedra such as $n$-cubes have
exponentially many components in the spatial dimension $n$.) We plan
on tackling this problem by applying to IRVA the reduction techniques
proposed in~\cite{BD97}, which seems feasible thanks to the acyclicity
of their transition relation. This would improve substantially the
efficiency of the data structure for large spatial dimensions.

\section*{Acknowledgement}

We wish to thank Jérôme Leroux for fruitful technical discussions about
the data structure presented in this paper.

\bibliography{irva} 
\bibliographystyle{eptcs}

\end{document}